\documentclass[jphysa]{iopart}

\usepackage{graphicx}
\usepackage{amsfonts}
\usepackage{amssymb}
\usepackage{braket}
\usepackage{times,txfonts}
\usepackage{subfigure}
\usepackage{hyperref}

\begin{document}

\title[Unifying approach to the quantification of bipartite correlations by Bures distance]{\sf \bfseries Unifying approach to the quantification of bipartite correlations by Bures distance}

\author{Thomas R. Bromley$^{1}$,  Marco Cianciaruso$^{2,1}$, Rosario {Lo Franco}$^{3,4,1}$, and Gerardo Adesso$^{1}$}
\address{$^{1}$School of Mathematical Sciences, The University of Nottingham, University Park, Nottingham NG7 2RD, United Kingdom
\\$^{2}$Dipartimento di Fisica ``E. R. Caianiello'', Universit\`a degli Studi di Salerno, Via Ponte don Melillo, I-84084 Fisciano (SA), Italy
\\$^{3}$Dipartimento di Fisica e Chimica, Universit\`a di Palermo, Via Archirafi 36, I-90123 Palermo, Italy
\\$^{4}$Instituto de F\'isica de S\~{a}o Carlos, Universidade de S\~{a}o Paulo, P.O.~Box 369,  S\~{a}o Carlos, 13560-970 S\~{a}o Paulo, Brazil
}

\begin{abstract}
The notion of distance defined on the set of states of a composite quantum system can be used to quantify total, quantum and classical correlations in a unifying way. We provide new closed formulae for classical and total correlations of two-qubit Bell-diagonal states by considering the Bures distance. Complementing the known corresponding expressions for entanglement and more general quantum correlations, we thus complete the quantitative hierarchy of Bures correlations for Bell-diagonal states. We then explicitly calculate Bures correlations for two relevant families of states: Werner states and rank-2 Bell-diagonal states, highlighting the subadditivity which holds for total correlations with respect to the sum of classical and quantum ones when using Bures distance. Finally, we analyse a dynamical model of two independent qubits locally exposed to non-dissipative decoherence channels, where both quantum and classical correlations measured by Bures distance exhibit freezing phenomena, in analogy with other known quantifiers of correlations.
\end{abstract}


\section{Introduction}

Characterising the different forms of correlations shared by the constituents of a composite quantum system is essential for the theoretical understanding and for the operational exploitation of the quantum system itself \cite{Horodecki2009,Modi2012}. Correlations which cannot be amenable to a classical description, in particular, exhibit a rich variety in mixed states of bipartite and multipartite quantum systems. Nowadays, the notion of quantum correlations refers not only to entanglement, but to more general forms of correlations which are conventionally identified with the quantum discord \cite{Ollivier2001, Henderson2001}, and capture for instance the necessary disturbance induced on quantum states by any local measurement (which is nonzero even in all separable states apart from so-called classical-quantum states). Correspondingly, the portion of correlations which are left in the state after a minimally disturbing local measurement can be identified with the classical correlations originally shared by the subsystems.

Unlike entanglement, for which a resource theory is well established \cite{Horodecki2009}, proposals to quantify quantum and classical correlations in this more general paradigm are still relatively scarce, and the mathematical requirements that any such proposal has to obey to be regarded as a valid measure are still to be completely formalised \cite{Modi2012}. Yet an interesting phenomenology associated to these correlations is being uncovered in different physical contexts. For example, from a foundational perspective the sudden transition from a decay to a plateau regime for classical correlations between a quantum system and its measurement apparatus has been interpreted as characterising the finite-time emergence of the pointer basis in the apparatus~\cite{Cornelio2012,Paula2013}. Moreover, quantum correlations between noninteracting qubits have been shown to dynamically revive despite of decoherence thanks to memory effects of the local environment, independently of the quantum or classical nature of the environment~\cite{SaroD1,SaroD3,lofrancoreview,xulofranco2013NatComms}. Quantum correlations of ground states also appear to play an important role in the characterisation of exotic phases of quantum many-body systems~\cite{Jiang2012,Giampaolo2013,Marzolino2013}. More generally, from an operational viewpoint various forms of quantum correlations, including and beyond entanglement, can and typically do provide fundamental resources for quantum technologies~\cite{Horodecki2009,Modi2012}. Consequently, rigorously addressing the quantification of correlations is of paramount importance.

The notion of distance defined on the convex set of states of a quantum system paves the way for several geometric approaches to the quantification of correlations~\cite{Vedral1997,Modi2010,Dakic2010,LuoFu,Monras2011,Bellomo2012a,Nakano2013,tufodiscord,LQU,Roga2014}. We refer in particular to Refs.~\cite{Nakano2013,Roga2014} for a comparison among these approaches with respect to the quantification of quantum correlations. In this paper, we focus on bipartite systems and we follow the approach of Ref.~\cite{Modi2010}, according to which the minimum distance between a state $\rho$ and the set of states that do not possess a particular kind of correlations is a quantifier of that kind of correlations. Hence, the minimum distances between the state $\rho$ of a bipartite system and the sets of product, classical-quantum and separable states represent, respectively, the amount of total correlations, quantum correlations and entanglement of $\rho$.
Furthermore, the minimum distance between the set of closest classical-quantum states to $\rho$ and the set of product states represents the classical correlations of $\rho$.
This geometric approach to the quantification of correlations manifests several appealing features. First, it is unifying, thus allowing for a direct comparison among all the above mentioned notions of correlations \cite{Modi2010}. Second, it readily suggests generalisations to the multipartite setting~\cite{Blasone2008}.

In this paper we use specifically the Bures distance on the set of states to define geometric quantifiers of correlations. The Bures distance is defined as
\begin{equation}
D_{Bu}\left(\rho,\sigma \right) \equiv \sqrt{2\left(1-\sqrt{F(\rho,\sigma)}\right)},
\end{equation}
where $\rho$ and $\sigma$ are two arbitrary states while $F(\rho,\sigma)$ is the Uhlmann fidelity \cite{uhlmann}
\begin{equation}\label{uberman}
F(\rho,\sigma)\equiv{\left[\mathrm{Tr}\left(\sqrt{\sqrt{\rho}\sigma\sqrt{\rho}}\,\right) \right]}^2.
\end{equation}
The reason for choosing this distance instead of others stems from its quite peculiar, but desirable, properties. Bures distance is at the same time contractive under completely positive trace-preserving maps, locally Riemannian and its metric coincides with the quantum Fisher information~\cite{Braunstein1994,Petz1996,Sommers2003,Facchi2010}, thus playing a crucial role in high precision interferometry and quantum metrology. Moreover, the minimum Bures distance between a state $\rho$ and the set of classical-quantum states is simply related to the maximal success probability in the ambiguous quantum state discrimination of a family of states and prior probabilities depending on $\rho$ \cite{Spehner2013,Spehner2014}. The task of minimal error quantum state discrimination plays a fundamental role both in quantum communication and cryptography and has been realised experimentally using polarised light \cite{Huttner1996,Mohseni2004}. On the contrary, e.g., the Hilbert-Schmidt distance is locally Riemannian but not contractive~\cite{Ozawa2000,Piani2012}, the trace distance is contractive but not locally Riemannian~\cite{Ruskai1994} and the relative entropy, although widely used in information theory, is technically not even a proper distance as it is not symmetric \cite{Modi2010}.

Here we derive closed formulae for classical and total correlations of Bell-diagonal states of two qubits according to the Bures distance. Together with the known corresponding formulae for entanglement~\cite{Streltsov2010} and discord-type quantum correlations~\cite{Aaronson2013a,Spehner2013,Spehner2014}, these allow us to gain a complete and unifying view of Bures correlations for Bell-diagonal states. We then provide two applications of these results. We first report the explicit expressions of the Bures correlations for two special subclasses of Bell-diagonal states, namely Werner states and rank-2 Bell-diagonal states. Finally, we consider a dynamical system made of two independent qubits locally interacting with a bosonic non-dissipative channel and show that both quantum and classical correlations measured by Bures distance can alternatively freeze during the evolution, joining the ranks of other faithful correlation quantifiers \cite{Mazzola,Aaronson2013a,Paula2013}. It is worthwhile to note that the freezing analysis was addressed in Ref.~\cite{Aaronson2013} with similar methods for the trace distance, but with a different definition of classical correlations.

The paper is organised as follows. In Section \ref{sec:quantumcorrelations} we review some known results concerning Bures quantum correlations and entanglement of Bell-diagonal states. In Sections \ref{sec:classicalcorrelations} and \ref{sec:totalcorrelations} we provide, respectively, the closed formulae for Bures classical and total correlations of Bell-diagonal states. In Section \ref{sec:examples} we compute the correlations of two particular classes of Bell-diagonal states, i.e., Werner states and rank-2 Bell-diagonal states. In Section \ref{sec:dynamics} we analyse the dynamics of correlations between two noninteracting qubits initially prepared in a Bell-diagonal state and subject to identical local pure dephasing channels. We conclude in Section \ref{sec:conclusions} with a summary and outlook.

\section{Quantum Correlations}\label{sec:quantumcorrelations}
Quantum correlations stem from two peculiar ingredients of quantum mechanics, the superposition principle and the tensor product structure of the Hilbert space associated to a composite quantum system. They are completely characterised by entanglement in the case of pure states, whereas in the case of mixed states entanglement constitutes only a part of the quantumness of correlations~\cite{Ollivier2001, Henderson2001}. As a result, for any pair of comparable quantifiers of general quantum correlations and entanglement, the quantum correlations of a state should intuitively be always greater or equal to the corresponding entanglement, being equal if the state is pure \cite{Interplay,PianiAdesso}. This is nicely captured by the aforementioned geometric approach. Specifically, in this paper, the quantum correlations of a state $\rho$ are quantified by the minimum Bures distance of $\rho$ to the set of classical-quantum states, namely
\begin{equation}
Q_{Bu}\left( \rho\right) \equiv \inf_{\chi\in\mathcal{CQ}} D_{Bu}\left(\rho,\chi \right) = D_{Bu}\left(\rho,\chi_\rho \right),
\end{equation}
where $\mathcal{CQ}$ is the set of classical-quantum states, i.e. states of the form $\chi=\sum_i p_i |i^A\rangle\langle i^A|\otimes\rho_i^B$ with ${\left\lbrace p_i \right\rbrace}$ being a probability vector, $\left\lbrace|i^A\rangle \right\rbrace $ an orthonormal basis of qubit $A$ and $\rho_i^B$ any state of qubit $B$ and $\chi_\rho$ is any of the closest classical-quantum states to $\rho$. The entanglement of $\rho$ is measured by the minimum Bures distance of $\rho$ to the set of separable states, namely
\begin{equation}
E_{Bu}\left( \rho\right) \equiv \inf_{\sigma\in\mathcal{S}} D_{Bu}\left(\rho,\sigma \right) = D_{Bu}\left(\rho,\sigma_\rho \right),
\end{equation}
where $\mathcal{S}$ is the set of separable states, i.e. states of the form $\sigma=\sum_i p_i \rho_i^A\otimes\rho_i^B$ with ${\left\lbrace p_i \right\rbrace}$ being a probability vector, $\rho_i^A$ and $\rho_i^B$ any state of qubit $A$ and $B$, respectively, while $\sigma_\rho$ is any of the nearest separable states to $\rho$. As the set of classical-quantum states is contained in the set of separable states, we immediately have that $Q_{Bu}\left( \rho\right)\geq E_{Bu}\left( \rho\right)$ for every $\rho$.

We shall restrict ourselves to the relevant but structurally simple class of Bell-diagonal (BD) states $\rho$ of two qubits, which are diagonal in the ``magic basis'' of the four maximally entangled Bell states. As a result, BD states are represented in the standard computational basis by the following matrix
\begin{equation}
\rho = \frac{1}{4}\left(\mathbb{I}^A\otimes\mathbb{I}^B + \sum_{i=1}^{3} c_i \sigma_i^A\otimes\sigma_i^B \right),
\end{equation}
where $\mathbb{I}$ and $\sigma_i$, $i=1, 2, 3$, are the identity and the Pauli matrices, respectively. The coefficients $c_i=\mathrm{Tr}\left[\rho\left(  \sigma_i^A\otimes\sigma_i^B\right)  \right] $ are the only correlation matrix elements of a BD state $\rho$ that can be different from zero, in terms of which the eigenvalues of $\rho$ are expressed as follows,
\begin{eqnarray}
\alpha &=& \frac{1}{4}\left(1+c_1-c_2+c_3 \right), \\
\beta &=& \frac{1}{4}\left(1-c_1+c_2+c_3 \right), \nonumber\\
\gamma &=& \frac{1}{4}\left(1+c_1+c_2-c_3 \right), \nonumber\\
\delta &=& \frac{1}{4}\left(1-c_1-c_2-c_3 \right). \nonumber
\end{eqnarray}
BD states are also called states with maximally mixed marginals, due to the fact that their reduced density matrices $\rho_A=\mathrm{Tr}_B\left( \rho\right) $ and $\rho_B=\mathrm{Tr}_A\left(  \rho\right)$ are both equal to the maximally mixed state of a qubit, i.e., $\rho_A=\frac{1}{2}\mathbb{I}^A$ and $\rho_B=\frac{1}{2}\mathbb{I}^B$. The class of BD states is particularly interesting: for instance, they include the well-known Bell states and Werner states \cite{Horodecki2009} and constitute a resource for entanglement activation and distribution \cite{PianiAdesso,Sciarrino,kay2012,Fedrizzi2013}.

In Ref.~\cite{Aaronson2013a,Spehner2013,Spehner2014} it was proven that, according to the Bures distance, one of closest classical-quantum states to a BD state $\rho$ is always a BD classical-quantum state of the form
\begin{equation}\label{Eq:BDCQState}
\chi_\rho^{BD} = \frac{1}{4}\left(\mathbb{I}^A\otimes\mathbb{I}^B +  s_k \sigma_k^A\otimes\sigma_k^B \right),
\end{equation}
where the index $k$ is such that $\Lambda_k =\Lambda_{max}\equiv\max\left\lbrace\Lambda_1,\Lambda_2,\Lambda_3\right\rbrace$, with
\begin{eqnarray}\label{Lambdas}
\Lambda_1 &\equiv& \sqrt{\alpha\gamma} + \sqrt{\beta\delta}, \nonumber\\
\Lambda_2 &\equiv& \sqrt{\alpha\delta} + \sqrt{\beta\gamma}, \\
\Lambda_3 &\equiv& \sqrt{\alpha\beta} + \sqrt{\gamma\delta},\nonumber
\end{eqnarray}
and
\begin{eqnarray}\label{Eq:theparameters}
s_1 &\equiv& \frac{\alpha+\gamma-\beta-\delta+2\left(\sqrt{\alpha\gamma} - \sqrt{\beta\delta} \right) }{1+2\Lambda_{1}}, \nonumber\\
s_2 &\equiv&  \frac{\beta+\gamma-\alpha-\delta+2\left(\sqrt{\beta\gamma}-\sqrt{\alpha\delta} \right) }{1+2\Lambda_{2}}, \\
s_3 &\equiv&  \frac{\alpha+\beta-\gamma-\delta+2\left(\sqrt{\alpha\beta} - \sqrt{\gamma\delta} \right) }{1+2\Lambda_{3}}.\nonumber
\end{eqnarray}
As a result, the closed expression for the quantum correlations of an arbitrary BD state, as quantified by the Bures distance, is given by
\begin{equation}\label{Buresdiscord}
Q_{Bu}\left( \rho\right) = \sqrt{2\left(1-\sqrt{ F(\rho,\chi_\rho^{BD})}\right)},
\end{equation}
where
\begin{equation}
F(\rho,\chi_\rho^{BD})=\frac{1}{2}\left(1+2 \Lambda_{max} \right) .
\end{equation}
We now make some important remarks. The index $k$ characterising the state $\chi_\rho^{BD}$ in (\ref{Eq:BDCQState}) is such that $|c_k|=\max\{|c_1|,|c_2|,|c_3|\}$. This can be easily proven by considering the expressions of the correlation matrix coefficients $c_i$ in terms of the eigenvalues of the BD state $\rho$ and noting that the condition $c_k^2=\max\{c_1^2,c_2^2,c_3^2\}$ is equivalent to the condition $\Lambda_k^2=\max\{\Lambda_1^2,\Lambda_2^2,\Lambda_3^2\}$.

The closest classical-quantum state $\chi_{\rho}$ to a BD state $\rho$ is unique if, and only if, $\rho$ is within the interior of the tetrahedron of BD states ($\alpha,\beta,\gamma,\delta > 0$) and the index $k$ such that $|c_k|=\max\{|c_1|,|c_2|,|c_3|\}$ is unique. Otherwise, there are infinitely many closest classical-quantum states to a BD state $\rho$ \cite{Spehner2014}.

We finally note that the Bures quantum correlations of $\rho$ as captured by Eq.~(\ref{Buresdiscord}) are different (conceptually and quantitatively) from the  ``discord of response'' of $\rho$, where quantumness of correlations is alternatively defined in terms of the minimum (Bures) distance between $\rho$ and the set of states obtained by rotating $\rho$ via local root-of-unity unitary operations on one subsystem only \cite{Roga2014}.

The closed expression for the entanglement of an arbitrary two-qubit state $\rho$, as quantified by the Bures distance, was obtained in terms of the concurrence \cite{Horodecki2009} $\mathrm{Con}(\rho)$ of $\rho$ and is given by \cite{Streltsov2010}
\begin{equation}
E_{Bu}\left( \rho\right) =  \sqrt{2\left(1-\sqrt{F(\rho,\sigma_\rho)}\right)},
\end{equation}
where
\begin{equation}
F(\rho,\sigma_\rho) =\frac{1}{2} \left(1+\sqrt{1-\mathrm{Con}^2(\rho)} \right).
\end{equation}
In the case of a BD state $\rho$, the concurrence specialises to
\begin{equation}
\mathrm{Con}(\rho)=\max\left\lbrace 0, \lambda_1 -\lambda_2 - \lambda_3 - \lambda_4  \right\rbrace
\end{equation}
with $\lambda_1\geq\lambda_2\geq \lambda_3\geq\lambda_4$ being the eigenvalues $\alpha,\beta,\gamma,\delta$ in non-increasing order.

\section{Classical Correlations}\label{sec:classicalcorrelations}
The classical correlations of a state $\rho$ can be quantified as follows. Given the set of all the closest classical-quantum states to $\rho$, called $\mathcal{CCQ}_{\rho}$, we define the classical correlations of $\rho$ to be
\begin{equation}\label{ClassicalCorrelationsDefinition}
C_{Bu}\left( \rho\right) \equiv \inf_{\chi_{\rho} \in \mathcal{CCQ}_{\rho}} \inf_{\pi\in\mathcal{P}} D_{Bu}\left(\chi_\rho,\pi \right) = \inf_{\chi_{\rho} \in \mathcal{CCQ}_{\rho}} D_{Bu}\left(\chi_\rho,\pi_{\chi_\rho}\right),
\end{equation}
where $\mathcal{P}$ is the set of  product states, i.e.~states of the form $\pi=\rho^A\otimes\rho^B$ with $\rho^A$ ($\rho^B$) being an arbitrary state of qubit $A$ ($B$) while $\pi_{\chi_\rho}$ is any of the closest product states to $\chi_\rho$. Notice that the definition in Eq.~(\ref{ClassicalCorrelationsDefinition}) represents an important improvement over previous attempts to quantify classical correlations geometrically  \cite{Modi2010,Aaronson2013,PaulaEPL}. In fact, without the inclusion of the additional minimisation over all the closest classical-quantum states to $\rho$, a measure of classical correlations might be ill-defined, as the distances between each closest classical-quantum state and their respective closest product states can generally differ. This issue has been very recently highlighted in an independent work \cite{Sarandy00}.

As we have already mentioned, if $\rho$ is an arbitrary BD state then, within $\mathcal{CCQ}_{\rho}$, there always exists a BD classical-quantum state $\chi_{\rho}^{BD}$ of the form of Eq.~(\ref{Eq:BDCQState}). In the following we shall prove that, for any BD state $\rho$, the BD state $\chi_{\rho}^{BD}$ achieves the infimum over $\mathcal{CCQ}_{\rho}$ in Eq.~(\ref{ClassicalCorrelationsDefinition}) and that one of the product states $\pi_{\chi_\rho^{BD}}$ nearest to the BD classical-quantum state $\chi_\rho^{BD}$ is the tensor product of the marginals of a BD state $\rho$, i.e.
\begin{equation}
\pi_{\chi_\rho^{BD}} = \frac{1}{4}\mathbb{I}^A\otimes\mathbb{I}^B.
\end{equation}
Thus, the Bures classical correlations of any BD state $\rho$ are quantified by
\begin{equation}\label{BuresClassCorr}
C_{Bu}\left( \rho\right) = \sqrt{2\left(1-\sqrt{ F(\chi_\rho^{BD},\pi_{\chi_\rho^{BD}})}\right)},
\end{equation}
with
\begin{equation}
F(\chi_\rho^{BD},\pi_{\chi_\rho^{BD}})=\frac{1+2(\Lambda_1+\Lambda_2+\Lambda_3) }{2(1+2 \Lambda_{max})},
\end{equation}
where $\Lambda_i$ ($i=1,2,3$) is defined in Eq. (\ref{Lambdas}).

We now prove the announced result.  In \cite{Spehner2014} the authors presented an explicit construction of all the closest classical-quantum states $\chi_\rho$ to any BD state $\rho$. In order to express $\chi_\rho$ in a general mathematical form we need to introduce some notations. Let $p_0=\delta$, $p_1=\beta$, $p_2=\alpha$, $p_3=\gamma$ be the eigenvalues of $\rho$ and $k$ be any index such that $|c_k|=\max\{|c_1|,|c_2|,|c_3|\}$. Let us finally introduce the orthonormal product basis $\{|\alpha_i,\beta_j\rangle=|\alpha_i\rangle\otimes|\beta_j\rangle\}_{i,j=0}^1$ of $\mathbb{C}^2\otimes\mathbb{C}^2$ defined as follows \cite{Spehner2014},
\begin{enumerate}
\item{if $k$ is unique, then $|\alpha_i\rangle=|\beta_i\rangle$ are the eigenvectors of $\sigma_k$};
\item{if $k$ can take two distinct values, then $k$, $|\alpha_i\rangle$ and $|\beta_i\rangle$ are defined by:
\begin{equation}
k=\cases{
1 & if $c_1=\pm c_2, |c_1|>|c_3|$, \\
3 & if $c_1=\pm c_3, |c_1|>|c_2|$,\\
3 & if $c_2=\pm c_3, |c_2|>|c_1|$,}
\end{equation}
\begin{equation}
|\alpha_i\rangle=\cases{
e^{-i\frac{\phi}{2}\sigma_3}\frac{|0\rangle+(-1)^i|1\rangle}{\sqrt{2}} & if $c_1=\pm c_2, |c_1|>|c_3|$, \\
e^{-i\frac{\theta}{2}\sigma_2}|i\rangle & if $c_1=\pm c_3, |c_1|>|c_2|$,\\
e^{i\frac{\theta}{2}\sigma_1}|i\rangle  & if $c_2=\pm c_3, |c_2|>|c_1|$,}
\end{equation}
and
\begin{equation}
|\beta_i\rangle=\cases{
e^{\mp i\frac{\phi}{2}\sigma_3}\frac{|0\rangle+(-1)^i|1\rangle}{\sqrt{2}} & if $c_1=\pm c_2, |c_1|>|c_3|$, \\
e^{\mp i\frac{\theta}{2}\sigma_2}|i\rangle & if $c_1=\pm c_3, |c_1|>|c_2|$,\\
e^{\pm i\frac{\theta}{2}\sigma_1}|i\rangle  & if $c_2=\pm c_3, |c_2|>|c_1|$;}
\end{equation}
}
where $\phi\in[0,2\pi[$ and $\theta\in[0,2\pi[$ are arbitrary;
\item{if $k$ can take all three different values, i.e. $c_1=\epsilon_2 c_2 = \epsilon_3 c_3$ with $\epsilon_{2,3}\in\{-1,1\}$, then we define
\begin{eqnarray}
k=3, \\
|\alpha_i\rangle=e^{-i\frac{\phi}{2}\sigma_3}e^{-i\frac{\theta}{2}\sigma_2}|i\rangle, \\
|\beta_i\rangle=e^{-i\epsilon_2\frac{\phi}{2}\sigma_3}e^{-i\epsilon_3\frac{\theta}{2}\sigma_2}|i\rangle ;
\end{eqnarray}
}
\end{enumerate}
where $\phi\in[0,2\pi[$ and $\theta\in[0,2\pi[$ are arbitrary.

 Now we are ready to write down the general form of any closest classical-quantum states $\chi_\rho$ to an arbitrary BD state $\rho$ \cite{Spehner2014}:
\begin{enumerate}

\item{\label{enum:probcond1} if $p_0p_k=0$ and $p_i p_j>0$ for $i\neq j$,  $i\neq k$, $j\neq k$, then
\begin{eqnarray}\label{Eq:generalformofchirhoprobcond1}
\chi_\rho(r)=\frac{1+s_k}{4}\left[|\alpha_0,\beta_0\rangle\langle\alpha_0,\beta_0|+ |\alpha_1,\beta_1\rangle\langle\alpha_1,\beta_1| \right] \\ \nonumber
+ \frac{1-s_k}{4}\left[(1+r)|\alpha_0,\beta_1\rangle\langle\alpha_0,\beta_1|+(1-r)|\alpha_1,\beta_0\rangle\langle\alpha_1,\beta_0|  \right],
\end{eqnarray}
where $r$ is a parameter which can take any value in the interval $r\in[-1,1]$ ;}
\item{\label{enum:probcond2} if $p_0p_k>0$ and $p_i p_j=0$  for $i\neq j$,  $i\neq k$, $j\neq k$, then
\begin{eqnarray}\label{Eq:generalformofchirhoprobcond2}
\chi_\rho(r)=\frac{1+s_k}{4}\left[(1+r)|\alpha_0,\beta_0\rangle\langle\alpha_0,\beta_0|+ (1-r)|\alpha_1,\beta_1\rangle\langle\alpha_1,\beta_1| \right] \\ \nonumber
+ \frac{1-s_k}{4}\left[|\alpha_0,\beta_1\rangle\langle\alpha_0,\beta_1|+|\alpha_1,\beta_0\rangle\langle\alpha_1,\beta_0|  \right],
\end{eqnarray}
where $r$ is a parameter which can take any value in the interval $r\in[-1,1]$ ;}
\item{\label{enum:probcond3} if $p_0p_1 p_2 p_3>0$, then
\begin{eqnarray}\label{Eq:generalformofchirhoprobcond3}
\chi_\rho=\frac{1+s_k}{4}\left[|\alpha_0,\beta_0\rangle\langle\alpha_0,\beta_0|+ |\alpha_1,\beta_1\rangle\langle\alpha_1,\beta_1| \right] \\ \nonumber
+  \frac{1-s_k}{4}\left[|\alpha_0,\beta_1\rangle\langle\alpha_0,\beta_1|+|\alpha_1,\beta_0\rangle\langle\alpha_1,\beta_0|  \right];
\end{eqnarray}}
\end{enumerate}
where $s_k$ is given by Eq. (\ref{Eq:theparameters}). Also, in the following it will be useful to note that $p_0p_k=0$ and $p_i p_j>0$ for $i\neq j$,  $i\neq k$, $j\neq k$, imply $s_k\in[\frac{3}{5},1]$, whereas  $p_0p_k>0$ and $p_i p_j=0$  for $i\neq j$,  $i\neq k$, $j\neq k$ imply $s_k\in[-1,-\frac{3}{5}]$, as can be easily seen from Eq. (\ref{Eq:theparameters}).

Now, for the sake of simplicity, let us focus on BD states $\rho$ such that $k$ is unique and equal to $3$ and such that $p_0p_3=0$ and $p_1p_2>0$, which from now on will be referred to as reference BD states. Later, we will generalise the analysis valid for this particular reference BD states, to a general BD state.  In the case of the reference BD states we have that the set of all closest classical-quantum states to $\rho$ is given by the following $1$-parameter family of states:
\begin{eqnarray}\label{Eq:CCQstatestoparticularBDstates}
\chi_\rho(r)=\frac{1+s_3}{4}\left[|00\rangle\langle 00|+ |11\rangle\langle11| \right] \\ \nonumber
+ \frac{1-s_3}{4}\left[(1+r)|01\rangle\langle 01|+(1-r)|10\rangle\langle10|  \right],
\end{eqnarray}
where $r\in[-1,1]$ and $s_3$ is given by Eq.~(\ref{Eq:theparameters}). In particular, for $r=0$ we get the BD classical-quantum state  $\chi_{\rho}^{BD}$ of the form of Eq.~(\ref{Eq:BDCQState}). Also, as we have already mentioned, due to the conditions $k=3$, $p_0p_3=0$ and $p_1p_2>0$, we have necessarily  $s_3\in[\frac{3}{5},1]$.

Let us consider a general product state of two qubits, $\pi=\rho^A\otimes\rho^B$, where $\rho^A$ and $\rho^B$ are any two states of qubits $A$ and $B$, respectively. Due to the Bloch representations of $\rho^A$ and $\rho^B$, i.e. $\rho^A=\frac{1}{2}\left(\mathbb{I}^A + \sum_{i=1}^3 a_i \sigma_i^A \right) $ and $\rho^B=\frac{1}{2}\left(\mathbb{I}^B + \sum_{i=1}^3 b_i \sigma_i^B \right)$ with $\left|\vec{a}\right|\leqslant 1,\left|\vec{b}\right|\leqslant 1$ , we have that $\pi$ is represented in the standard computational basis by the following matrix
\begin{equation}\label{statoprodotto}\pi = \frac{1}{4}\left(\mathbb{I}^A\otimes\mathbb{I}^B +  \sum_{i=1}^3 a_i \sigma_i^A\otimes\mathbb{I}^B  + \sum_{i=1}^3 b_i\mathbb{I}^A\otimes\sigma_i^B + \sum_{i,j=1}^3 a_i b_j \sigma_i^A\otimes\sigma_j^B \right ).\end{equation}
A general product state $\pi$ of two qubits is clearly characterised by the Bloch vectors of each qubit, i.e. $\vec{a}\equiv\left\lbrace a_1,a_2,a_3 \right\rbrace $ and $\vec{b}\equiv\left\lbrace b_1,b_2,b_3 \right\rbrace $. We now take into account the product states $\pi'$ and $\pi_0$, where  $\pi'=U_3 \pi U_3^\dagger$, with $U_3=\sigma_3^A\otimes\mathbb{I}^B$, is characterised by the Bloch vectors $\vec{a}'=\{-a_1,-a_2,a_3\}$ and   $\vec{b}'=\vec{b}=\{b_1,b_2,b_3\}$, whereas $\pi_0\equiv \frac{1}{2}\left(\pi + \pi' \right) $ is characterised by the Bloch vectors $\vec{a}_0=\{0,0,a_3\}$ and   $\vec{b}_0=\vec{b}=\{b_1,b_2,b_3\}$. Then, the following holds
\begin{equation}\label{Eq:FidelityEquality}
F\left(\chi_\rho,\pi \right)=F\left(\chi_\rho,\pi' \right),
\end{equation}
 where $\chi_\rho$ is any closest classical-quantum state of the form (\ref{Eq:CCQstatestoparticularBDstates}). To prove the above equality it suffices to consider the invariance of the fidelity under general unitaries and the invariance of any $\chi_\rho$ under the action of the particular local unitary $U_3$.

It is known that the fidelity is a concave function on the convex set of states, i.e.
\begin{equation}\label{eq:concavityfidelity}
F(\rho,p\sigma_1 + (1-p)\sigma_2)\geq pF(\rho,\sigma_1) + (1-p)F(\rho,\sigma_2),\ \ \forall p\in\left[0,1 \right]
\end{equation}
for any states $\rho$, $\sigma_1$ and $\sigma_2$. As a result, by substituting $p=\frac{1}{2}$, $\rho=\chi_\rho$, $\sigma_1 = \pi$ and $\sigma_{2} =\pi'$ into (\ref{eq:concavityfidelity}), one obtains
\begin{equation}
F\left(\chi_\rho,\pi_0 \right)\geq F\left(\chi_\rho,\pi \right).
\end{equation}
By symmetry, a similar result holds also by flipping the first two components of the Bloch vector $\vec{b}$.
As a result, in order to maximise the fidelity between any closest classical-quantum state $\chi_\rho$ to a reference BD state $\rho$ and any product state $\pi$, the Bloch vectors that characterise $\pi$ must be necessarily such that $a_i=a \delta_{i3}$ and $b_i=b \delta_{i3}$.
The square root of the fidelity between any $\chi_\rho$ of the form (\ref{Eq:CCQstatestoparticularBDstates}) and any product state $\pi$ with $a_i=a \delta_{i3}$ and $b_i=b \delta_{i3}$ is
\begin{eqnarray}\label{Eq:ClassicalFidelity}
\sqrt{F(\chi_{\rho},\pi)}= \nonumber\\
\frac{1}{4} \sqrt{(1+a) (1-b)(1-r) (1-s_{3})}+\frac{1}{4} \sqrt{(1-a) (1+b) (1+r)(1-s_{3})} \nonumber\\
+\frac{1}{4} \sqrt{(1-a) (1-b) (1+s_{3})}+\frac{1}{4} \sqrt{(1+a) (1+b) (1+s_{3})},
\end{eqnarray}
where we have used the fact that any $\chi_{\rho}$ of the form (\ref{Eq:CCQstatestoparticularBDstates}) commutes with any product state $\pi$ with $a_i=a \delta_{i3}$ and $b_i=b \delta_{i3}$, so that the square root of their fidelity is nothing but the square root of their classical fidelity which, in turn, is given by the sum of the square roots of the products of the corresponding eigenvalues of $\chi_{\rho}$ and $\pi$.
By maximising Eq. (\ref{Eq:ClassicalFidelity}) with respect to $r$, $a$ and $b$, one obtains that $r=a=b=0$ reaches the global maximum for any $s_3\in[0, 1]$, so in particular for any $s_3\in[\frac{3}{5}, 1]$. As a consequence, for any reference BD state $\rho$, the infimum over $\mathcal{CCQ}_\rho$  in Eq. (\ref{ClassicalCorrelationsDefinition}) is achieved by $\chi_\rho^{BD}$  and one of the nearest product states to $\chi_\rho^{BD}$ is $\pi_{\chi_\rho^{BD}}= \frac{1}{4}\mathbb{I}^A\otimes\mathbb{I}^B$.

Before generalising the above analysis from the reference BD states to any BD state we need to make the following two remarks. First, for any orthonormal product basis $\{|\alpha_i,\beta_j\rangle\}_{i,j=0}^1$, any classical-quantum state in Eq.~(\ref{Eq:generalformofchirhoprobcond1})  can be transformed through a local unitary into the reference classical-quantum state in Eq.~(\ref{Eq:CCQstatestoparticularBDstates})  with the same value of $r$ and $s_3=s_k$, any classical-quantum state in Eq.~(\ref{Eq:generalformofchirhoprobcond2}) can be transformed through a local unitary into the reference classical-quantum state  (\ref{Eq:CCQstatestoparticularBDstates}) with the same value of $r$ and $s_3=-s_k$, and finally any classical-quantum state in Eq.~(\ref{Eq:generalformofchirhoprobcond3}) can be transformed through a local unitary into the reference classical-quantum state  (\ref{Eq:CCQstatestoparticularBDstates}) with $r=0$ and $s_3=|s_k|$. Second, we note that the minimal Bures distance from the set of product states is invariant under local unitaries, indeed
\begin{eqnarray}\label{Eq:BuresLocalUnitaries}
\inf_{\pi \in \mathcal{P}} D\left(\chi_{\rho},\pi\right) &=\inf_{\pi \in \mathcal{P}} D\left( (U_{A} \otimes U_{B})\chi_{\rho}(U_{A}^{\dagger} \otimes U_{B}^{\dagger}), (U_{A} \otimes U_{B})\pi(U_{A}^{\dagger} \otimes U_{B}^{\dagger})\right) \nonumber \\
 &=\inf_{\pi \in \mathcal{P}} D\left( (U_{A} \otimes U_{B})\chi_{\rho}(U_{A}^{\dagger} \otimes U_{B}^{\dagger}),\pi\right)
\end{eqnarray}
where in the first equality we use the invariance of the Bures distance under unitaries and in the second equality we use the locality and bijectivity of local unitaries.

As promised, we are now ready to generalise the above analysis from the reference BD states to any BD state. Let us start from the BD states $\rho$ satisfying Condition (\ref{enum:probcond1}), i.e. the ones such that $p_0p_k=0$ and $p_i p_j>0$  for $i\neq j$,  $i\neq k$, $j\neq k$. As we mentioned, we have necessarily a reference BD state $\rho'$ with $s_3=s_k$ such that for any pair of $\chi_{\rho}(r) \in \mathcal{CCQ}_{\rho}$ and $\chi_{\rho'}(r) \in \mathcal{CCQ}_{\rho'}$ with the same value of $r$, there always exists a local unitary $U_{A} \otimes U_{B}$ such that $\chi_{\rho'} = (U_{A} \otimes U_{B}) \chi_{\rho} (U_{A}^{\dagger} \otimes U_{B}^{\dagger})$. Also,
\begin{eqnarray}\label{Eq:ChiRhoBDInfimum}
\inf_{\chi_\rho \in \mathcal{CCQ}_\rho}\inf_{\pi \in \mathcal{P}} D\left(\chi_{\rho},\pi\right) &= \min_{r\in[-1,1]}\inf_{\pi \in \mathcal{P}} D\left(\chi_{\rho}(r),\pi\right) \nonumber \\
&=  \min_{r\in[-1,1]}\inf_{\pi \in \mathcal{P}} D \left( \chi_{\rho'}(r),\pi\right) \nonumber \\
&=  D \left( \chi_{\rho'}^{BD},\frac{1}{4}\mathbb{I}^A\otimes\mathbb{I}^B\right) \nonumber \\
&=  D \left( \chi_{\rho}^{BD},\frac{1}{4}\mathbb{I}^A\otimes\mathbb{I}^B\right)
\end{eqnarray}
where in the first equality, we use the fact that all the states $\chi_\rho(r)$  with the same value of $r$, that may depend on $\theta$ or $\phi$ or both, are local unitarily equivalent and that the minimal Bures distance from the set of product states is invariant under local unitaries. In the second equality we use the fact that  all the states $\chi_\rho(r)$  are local unitarily equivalent to the states $\chi_{\rho'}(r)$ with the same value of $r$ and again that the minimal Bures distance from the set of product states is invariant under local unitaries. In the third equality we use the fact that $\chi_{\rho'}(r) $ achieves the infimum over $\mathcal{CCQ}_\rho$ is $\chi_{\rho'}(0) =\chi_{\rho'}^{BD}$ and one of its nearest product states is $\pi_{\chi_{\rho'}^{BD}}= \frac{1}{4}\mathbb{I}^A\otimes\mathbb{I}^B$. In the last equality we use the fact that $\chi_{\rho}^{BD}$ is local unitarily equivalent to $\chi_{\rho'}^{BD}$, both corresponding to $r=0$, and the fact that $\frac{1}{4}\mathbb{I}^A\otimes\mathbb{I}^B$ and the Bures distance are invariant under general unitaries. Equation (\ref{Eq:ChiRhoBDInfimum}) means that, for any BD state $\rho$ satisfying Condition (\ref{enum:probcond1}), i.e. such that $p_0p_k=0$ and $p_i p_j>0$  for $i\neq j$,  $i\neq k$, $j\neq k$, the infimum over $\mathcal{CCQ}_{\rho}$ in Eq.~(\ref{ClassicalCorrelationsDefinition}) is achieved by the BD closest classical-quantum states to $\rho$, $\chi_{\rho}^{BD}$, and one of the nearest product states to $\chi_\rho^{BD}$ is $\pi_{\chi_\rho^{BD}}= \frac{1}{4}\mathbb{I}^A\otimes\mathbb{I}^B$. An identical reasoning holds for BD states satisfying Condition (\ref{enum:probcond2}), i.e. such that $p_0p_k>0$ and $p_i p_j=0$  for $i\neq j$,  $i\neq k$, $j\neq k$, with the only exception of using a reference BD state $\rho'$ such that $s_3=-s_k$. Finally, for BD states satisfying Condition (\ref{enum:probcond3}), i.e. such that $p_0 p_1 p_2 p_3>0$, one simply needs to use the fact that each of the closest classical-quantum states  $\chi_\rho$, all obtained by setting $r=0$ in the previous two cases, is local unitarily equivalent to $\chi_{\rho'}^{BD}$ directly.

In summary, we have proven that for any BD state $\rho$, the closest BD classical-quantum state $\chi_{\rho}^{BD}$ and the product of the marginals of $\rho$ quite miraculously achieve the double minimisation in Eq.~(\ref{ClassicalCorrelationsDefinition}). This result is highly nontrivial (and not obvious {\it a priori}) and suggests that the definition of classical correlations proposed here is particularly natural for BD states $\rho$.

\section{Total Correlations}\label{sec:totalcorrelations}
The Bures total correlations of a state $\rho$ are defined by the minimum Bures distance of $\rho$ to the set of product states, namely
\begin{equation}
T_{Bu}\left( \rho\right) \equiv \inf_{\pi\in\mathcal{P}} D_{Bu}\left(\rho,\pi \right) = D_{Bu}\left(\rho,\pi_{\rho}\right),
\end{equation}
where $\mathcal{P}$ is the set of  product states, while $\pi_{\rho}$ is any of the product states closest to $\rho$. Therefore, in order to obtain the total correlations of a given BD state $\rho$, we simply need to maximise the fidelity $F(\rho,\pi)$ between $\rho$ and any product state $\pi$.

However, the argument used in the previous section to maximise the fidelity $F(\chi_\rho,\pi)$ between any closest classical-quantum state $\chi_\rho$ and any product state $\pi$, does not apply anymore to the present problem.  We note in fact that Eq.~(\ref{Eq:FidelityEquality}) does not hold for a general BD state, i.e.,
\begin{equation}
F(\rho,\pi) \neq F(\rho,\pi').
\end{equation}
The concavity of the fidelity thus cannot be utilised to extend the previous analysis from classical to total correlations.

We then formulate an ansatz on the form of one of the closest product states $\pi_{\rho}=\frac{1}{2}\left(\mathbb{I}^A + \sum_{i=1}^3 a_i \sigma_i^A \right)\otimes\frac{1}{2}\left(\mathbb{I}^B + \sum_{i=1}^3 b_i \sigma_i^B \right)$ to a general BD state $\rho$, which is of the form
\begin{equation}\label{eq:ansatzproductstates}
a_i=a \delta_{il},\ \ b_i=b \delta_{il},\ \ a=\frac{|c_l|}{c_l} b,
\end{equation}
where the index $l$ is such that $|c_l|=\min\{|c_1|,|c_2|,|c_3|\}$. This allows us to accomplish the optimisation of the Bures distance analytically. Interestingly, the ansatz form of one of the closest product states to a BD state using the trace distance, formulated in Ref. \cite{Aaronson2013}, is the same as Eq. (\ref{eq:ansatzproductstates}), but with $l$ given by $|c_l|=\max\{|c_1|,|c_2|,|c_3|\}$.

The ansatz in Eq. (\ref{eq:ansatzproductstates}) is supported and verified by an extensive numerical investigation, which was implemented in the following way. We begin by generating a random set of four normalised probabilities and forming a BD state by setting these probabilities as the eigenvalues $\alpha$, $\beta$, $\gamma$ and $\delta$. We then numerically maximise the fidelity between this random BD state and a general product state. The result of this numerical maximisation is compared with the analytical maximisation of the fidelity between the random BD state and our ansatz product states, Eq. (\ref{eq:ansatzproductstates}). This process was repeated for $10^6$ randomly generated BD states, and in all cases the analytically maximised fidelity between the random BD state and the ansatz product states exceeded or equalled the numerical maximisation over all product states.

In order to get the explicit expression of the above coefficient $a$ characterising the Bloch vectors of $\pi_\rho$, in terms of the BD state parameters, let us proceed as follows. We define the auxiliary set of coefficients $\vec{\mu} \equiv \{\mu_1,\mu_2,\mu_3,\mu_4\}$ as given by suitable reorderings of the BD state eigenvalues. Specifically, for $l=1$ we have $\vec{\mu}=\{\alpha,\gamma,\beta,\delta\}$, for $l=2$ we have $\vec{\mu}=\{\beta,\gamma,\alpha,\delta\}$, and for $l=3$ we have $\vec{\mu}=\{\alpha,\beta,\gamma,\delta\}$.

For any BD state $\rho$ with, respectively, $c_{l}\geq 0$ and $c_{l}\leq 0$, the square root of the fidelity between $\rho$ and any product state $\pi$ having the Bloch vectors of Eq.~(\ref{eq:ansatzproductstates}) is
\begin{eqnarray}
\sqrt{F(\rho,\pi)}= \nonumber \\
\frac{1}{2}\left[\sqrt{(\sqrt{\mu_1}+\sqrt{\mu_2})^2+(\sqrt{\mu_1}-\sqrt{\mu_2})^2a^2}+(\sqrt{\mu_3}+\sqrt{\mu_4})\sqrt{1-a^2}\right]
\end{eqnarray}
and
\begin{eqnarray}
\sqrt{F(\rho,\pi)}= \nonumber \\
\frac{1}{2}\left[\sqrt{(\sqrt{\mu_3}+\sqrt{\mu_4})^2+(\sqrt{\mu_3}-\sqrt{\mu_4})^2a^2}+(\sqrt{\mu_1}+\sqrt{\mu_2})\sqrt{1-a^2}\right].
\end{eqnarray}
By maximising now the square root of the fidelity between $\rho$ and $\pi$ with respect to $a$, one obtains after some algebra that:
\begin{enumerate}
\item{\label{Item:Condition1}if $c_l>0$ and the condition
\begin{equation}\label{eq:conditionfortherealityofaplus}
{\left( \sqrt{\mu_1}-\sqrt{\mu_2}\right)}^2>{\left( \sqrt{\mu_3}+\sqrt{\mu_4}\right)}{\left( \sqrt{\mu_1}+\sqrt{\mu_2}\right)}
\end{equation}
is fulfilled, then $b=a$ with
\begin{equation}\label{eq:optimalainthefirstcaseplus}
a=\pm\sqrt{\frac{{\left( \sqrt{\mu_1}-\sqrt{\mu_2}\right)}^4-{\left( \sqrt{\mu_3}+\sqrt{\mu_4}\right)}^2{\left( \sqrt{\mu_1}+\sqrt{\mu_2}\right)}^2 }{{\left( \sqrt{\mu_1}-\sqrt{\mu_2}\right)}^4+{\left( \sqrt{\mu_3}+\sqrt{\mu_4}\right)}^2{\left( \sqrt{\mu_1}-\sqrt{\mu_2}\right)}^2}};
\end{equation}}
\item{if $c_l<0$ and the condition
\begin{equation}\label{eq:conditionfortherealityofaminus}
{\left( \sqrt{\mu_3}-\sqrt{\mu_4}\right)}^2>{\left( \sqrt{\mu_1}+\sqrt{\mu_2}\right)}{\left( \sqrt{\mu_3}+\sqrt{\mu_4}\right)}
\end{equation}
is fulfilled, then $b=-a$ with
\begin{equation}\label{eq:optimalainthefirstcaseminus}
a=\pm\sqrt{\frac{{\left( \sqrt{\mu_3}-\sqrt{\mu_4}\right)}^4-{\left( \sqrt{\mu_1}+\sqrt{\mu_2}\right)}^2{\left( \sqrt{\mu_3}+\sqrt{\mu_4}\right)}^2 }{{\left( \sqrt{\mu_3}-\sqrt{\mu_4}\right)}^4+{\left( \sqrt{\mu_1}+\sqrt{\mu_2}\right)}^2{\left( \sqrt{\mu_3}-\sqrt{\mu_4}\right)}^2}};
\end{equation}}
\item{\label{Item:Condition3}if none of the two conditions above hold, we have $a=b=0$.}
\end{enumerate}
As a result, the Bures total correlations of an arbitrary BD state $\rho$ are measured by
\begin{equation}\label{Eq:TotalCorrelations}
T_{Bu} (\rho) = \sqrt{2 \left(1-\sqrt{F(\rho,\pi_{\rho})}\right)},
\end{equation}
where
\begin{equation}\label{eq:fidelitymax}
F(\rho,\pi_{\rho})=\cases{
F_+ & if (i) holds, \\
F_- & if (ii) holds,\\
F_{0} & if (iii) holds,}
\end{equation}
with
\begin{equation}\label{eq:fidelityplus}
F_+= \frac{(\mu_1 +\mu_2 ) \left(1 -2 \sqrt{\mu_1  \mu_2 } + 2 \sqrt{\mu_3  \mu_4 }\right)}{2 \left(\sqrt{\mu_1} - \sqrt{\mu_2} \right)^2},
\end{equation}
\begin{equation}\label{eq:fidelityminus}
F_-= \frac{(\mu_3 +\mu_4 ) \left(1 -2 \sqrt{\mu_3  \mu_4 } + 2 \sqrt{\mu_1  \mu_2 }\right)}{2 \left(\sqrt{\mu_3} - \sqrt{\mu_4} \right)^2},
\end{equation}
\begin{equation}
F_{0} = \frac{1}{4} \left( \sqrt{\alpha}+\sqrt{\beta}+\sqrt{\gamma}+\sqrt{\delta}\right) ^{2}.
\end{equation}

\begin{figure}[t]
    \centering
    \includegraphics[width=0.6\textwidth]{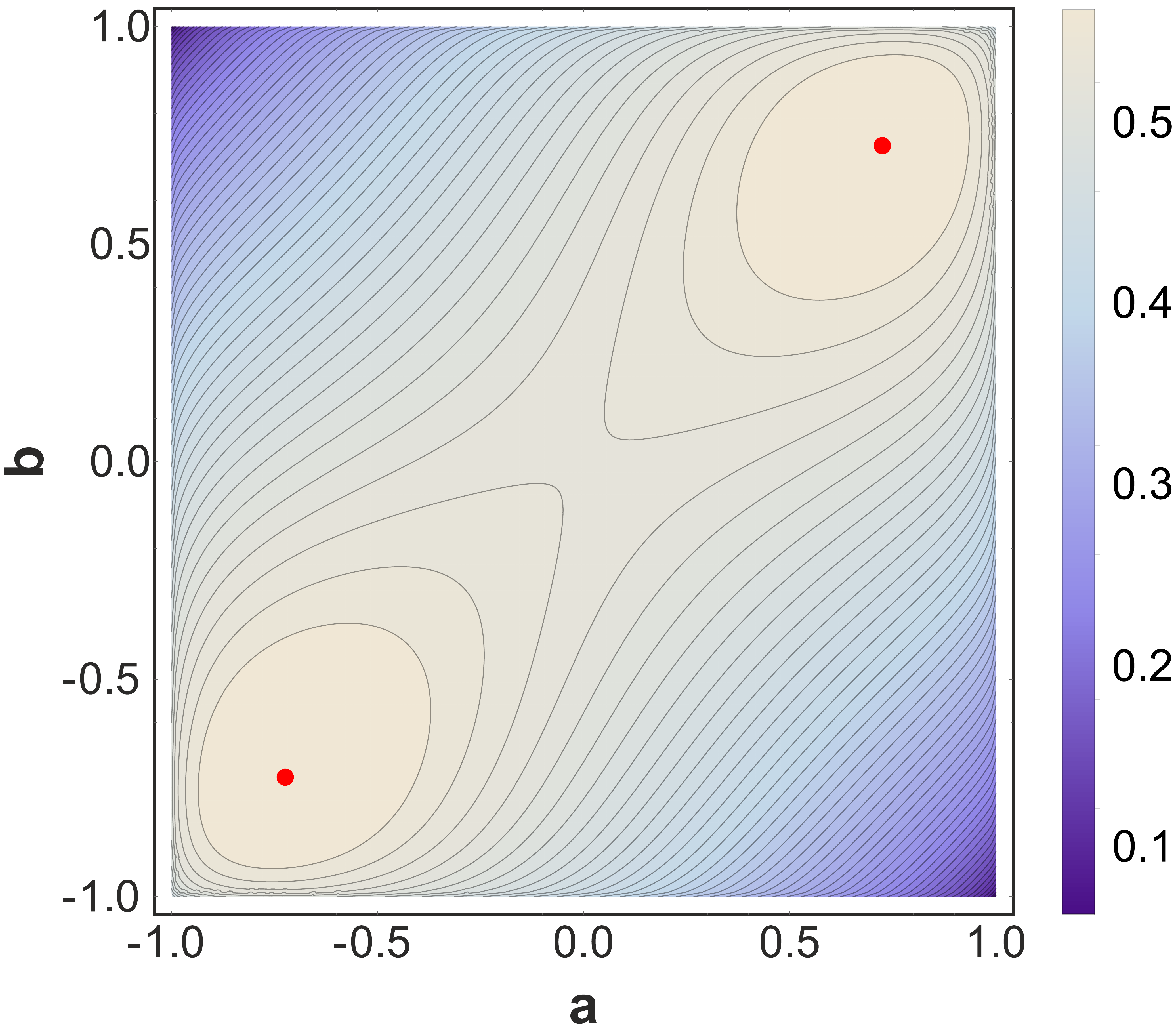}
    \caption{The fidelity between a BD state characterised by $\alpha = 0.874168$, $\beta=0.001239$, $\gamma=0.026908$, $\delta=0.097685$, obeying Condition (\ref{Item:Condition1}) with $l=3$, and product states obtained as perturbations of the ansatz from Eq.~(\ref{eq:ansatzproductstates}), by considering any value of $a$ and $b$ belonging to $[-1,1]$ and not only $a=\frac{|c_l|}{c_l} b$. The red dots indicate the maxima  found at $a=b= \pm 0.725398$, in agreement with the values predicted by Eq.~(\ref{eq:optimalainthefirstcaseplus}). It can be seen that the product of the marginals $\frac{1}{4} \mathbb{I}^{A} \otimes \mathbb{I}^{B}$ becomes a saddle point in the case of Condition (\ref{Item:Condition1}).}
    \label{Fig:Perturbations}
\end{figure}

The contour plot in Figure~\ref{Fig:Perturbations} shows the fidelity between an example BD state obeying Condition (\ref{Item:Condition1}) with $l=3$,
and product states obtained as perturbations of the ansatz from Eq.~(\ref{eq:ansatzproductstates}), by allowing $a$ and $b$ to vary over the interval $[-1,1]$. The values of $a$ and $b$ that maximise this fidelity are shown to coincide with the values given by Eq.~(\ref{eq:optimalainthefirstcaseplus}). Similar plots may be created by perturbing the ansatz states in such a way that the index $l$ is no longer $|c_l|=\min\{|c_1|,|c_2|,|c_3|\}$, i.e. $l=3$, but rather $l=1$ or $l=2$. These plots both show two maxima, but the fidelity in these two maxima never exceeds the maximal value corresponding to $l=3$.

It is worth highlighting that the product of the marginals does not represent, in general, the closest product state $\pi_{\rho}$ to an arbitrary BD state $\rho$; indeed, from the above classification, only when Condition (\ref{Item:Condition3}) holds the corresponding $F_0$ is then precisely the fidelity between $\rho$ and
$\frac{1}{4}\mathbb{I}^A\otimes\mathbb{I}^B$. This apparently counterintuitive feature has been observed as well when the trace distance is used to measure total correlations~\cite{Aaronson2013}. Differently, when Hilbert-Schmidt~\cite{Bellomo2012a} and relative entropy~\cite{Modi2010} distances are taken into account, the closest product state to any BD state is always the product of the marginals, i.e.~the maximally mixed state.

Another aspect to be remarked in this context is that, according to a naive intuition, total correlations may be expected to be equal to the sum of the classical and quantum correlations. Indeed, this is true for BD states when considering the relative entropy and Hilbert-Schmidt distance measures of correlations~\cite{Bellomo2012a}. However, the triangle inequality which is a requirement for any metric imposes that, for geometric quantifiers of correlations, only a subadditivity property needs to be satisfied, of the form $T \leq Q + C$. This turns out to be in general a sharp inequality when correlations are measured either by the trace distance~\cite{Aaronson2013,PaulaEPL} or by the Bures distance, as shall become apparent in the following.


\section{Examples}\label{sec:examples}
In this Section the Bures distance correlations are analysed for two families of one-parameter BD states: Werner states~\cite{Werner1989} and rank-2 BD states.

\begin{figure}[t]
    \centering
    \includegraphics[width=0.6\textwidth]{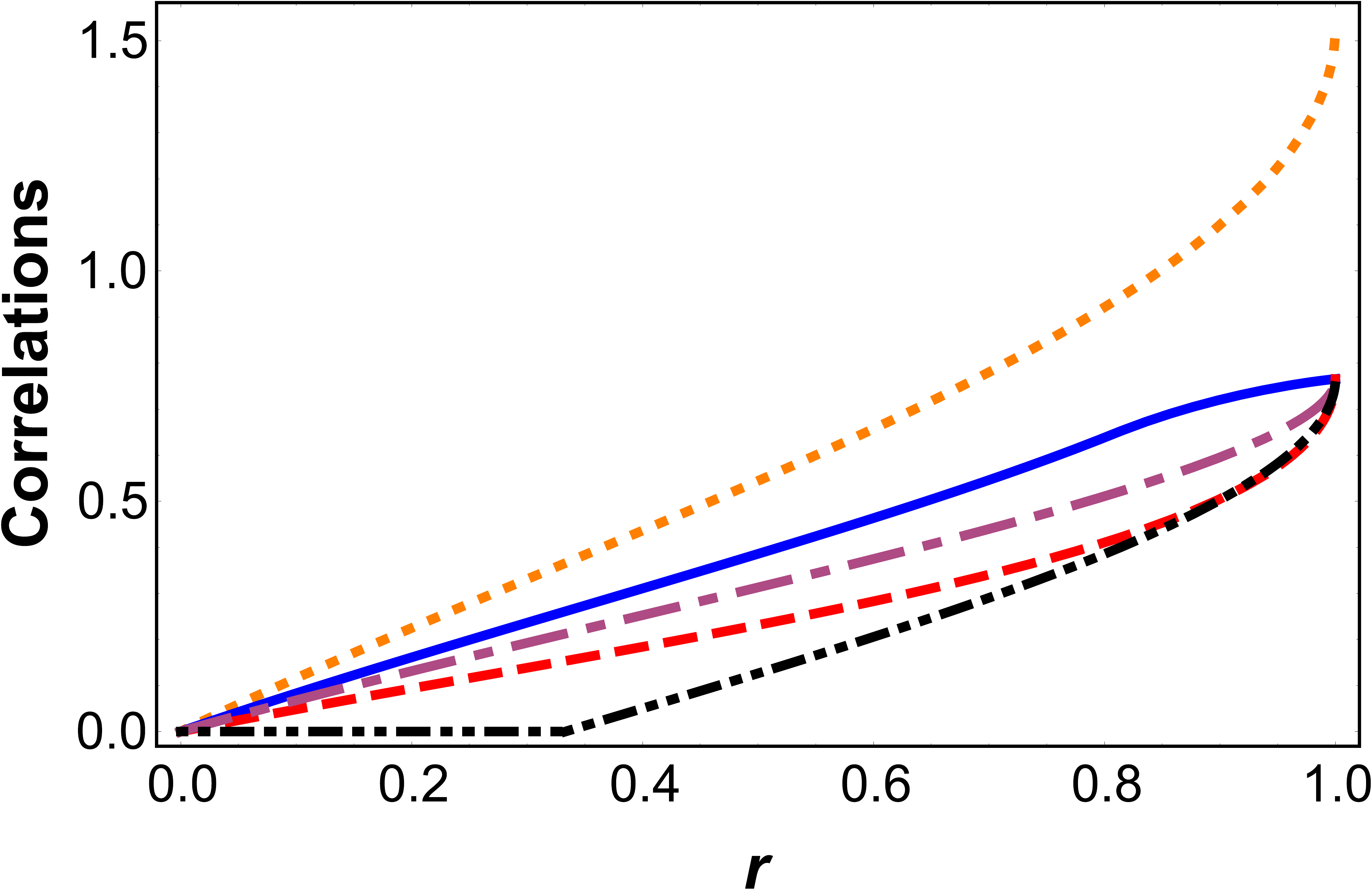}
    \caption{Bures distance based correlations for Werner states as a function of $r$. The figure displays total correlations (solid blue line), classical correlations (dashed red line), quantum correlations (dot-dashed purple line) and entanglement (dot-dot-dashed black line). The sum of classical and quantum correlations (dotted orange line) is also plotted.}
    \label{Fig:WernerCorrelations}
\end{figure}

Werner states $\rho_{W}$ are conventionally defined, for two qubits, as the mixture of a maximally entangled Bell state $\ket{\Phi} = (\ket{00} + \ket{11})/\sqrt{2}$ with the maximally mixed state, namely $\rho_{W} = r \ket{\Phi}\bra{\Phi} + (1-r)(\mathbb{I}^A\otimes\mathbb{I}^B)/4$, with $r \in [0,1]$.
Werner states are therefore a subclass of BD states, with eigenvalues given simply by $\alpha = (1+3r)/4$ and $\beta = \gamma =\delta = (1-r)/4$. Combining the results of \cite{Streltsov2010,Aaronson2013a,Spehner2013,Spehner2014} with the above analysis,  the Bures distance based correlations of $\rho_{W}$, as shown in Figure~\ref{Fig:WernerCorrelations} as functions of $r$, are given overall by 
\begin{equation}
E_{Bu}^{2}(\rho_{W})=\cases{
0 & $r\leq 1/3$,\\
2-\sqrt{2+\sqrt{3(1+2r-3r^2)}} & $r > 1/3$,}
\end{equation}
\begin{equation}
Q_{Bu}^{2}(\rho_{W}) = 2-\sqrt{3-r+\sqrt{1+2r-3r^2}},
\end{equation}
\begin{equation}
C_{Bu}^{2}(\rho_{W})=2 -\frac{ 3 \sqrt{1-r}+\sqrt{1+3r}}{\sqrt{3-r+\sqrt{1+2r-3r^2}}},
\end{equation}
\begin{equation}
T_{Bu}^{2}(\rho_{W})=\cases{
\Xi_{1}(r) & $r < (1 + \sqrt{5})/4$,\\
\Xi_{2}(r) & $r \geq (1 + \sqrt{5})/4$,} \qquad \mbox{with}
\end{equation}
\begin{equation*}
\Xi_{1}(r) = \frac{1}{2} \left(4-3 \sqrt{1-r}-\sqrt{3 r+1}\right),
\end{equation*}
\begin{equation*}
\Xi_{2}(r) = 2-\sqrt{\frac{(r+1) \left(3-r-\sqrt{1+2r-3 r^2}\right)}{1+r-\sqrt{1+2r-3 r^2}}}.
\end{equation*}
All the correlations vanish for $r=0$, when the state reduces to the maximally mixed (uncorrelated) state, and are monotonically increasing functions of $r$. However, entanglement and total correlations are not smooth functions of $r$, due to non-analyticities at $r=1/3$ and $r=(1 + \sqrt{5})/4$, respectively. For $r > (1 + \sqrt{5})/4$, the closest product state $\pi_{\rho}$ to $\rho$ is not the maximally mixed product of the marginals, and becomes instead increasingly less mixed with increasing $r$, eventually reaching a pure state (e.g.~$\ket{00}\bra{00}$) when $r=1$.  We finally note that for every $r$ we have $T_{Bu} \leq Q_{Bu}+C_{Bu} $, saturating the bound only trivially for $r=0$ where $Q_{Bu}=C_{Bu}=T_{Bu}=0$.


\begin{figure}[t]
    \centering
    \includegraphics[width=0.6\textwidth]{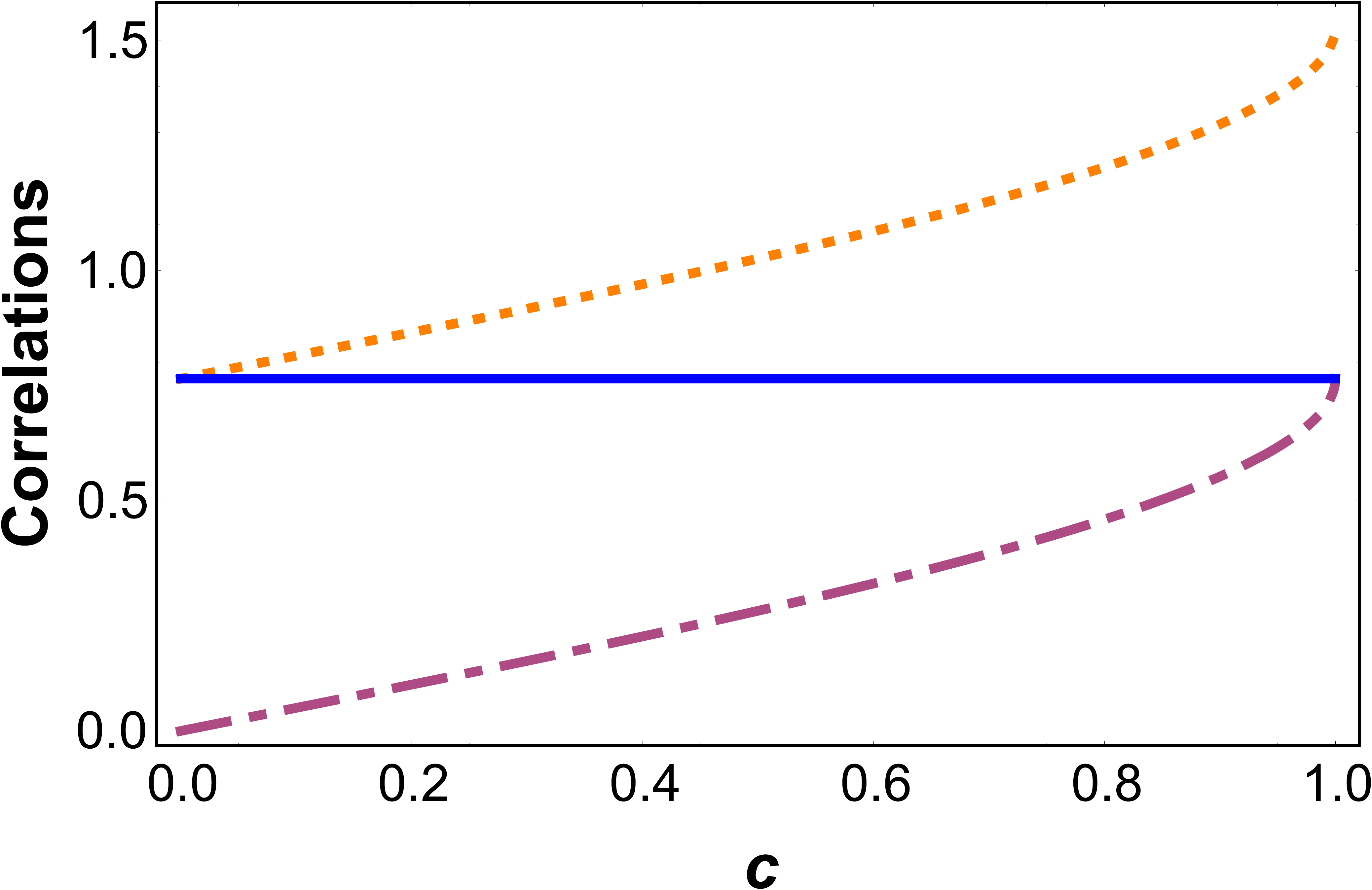}
    \caption{Bures distance correlations for rank-2 Bell-diagonal states as a function of $c$. The figure displays total correlations (solid blue line), equal to the classical correlations, and quantum correlations (dot-dashed purple line), which are equal to the entanglement. The sum of classical and quantum correlations (dotted orange line) is also plotted.}
    \label{Fig:Rank2Correlations}
\end{figure}

We next consider a class of rank-2 BD states $\rho_{2}$, whose eigenvalues take the form $\alpha=(1-c)/2$, $\beta=(1+c)/2$ and $\gamma=\delta=0$, with $c\in[0,1[$. The Bures distance correlations, shown in Figure~\ref{Fig:Rank2Correlations} as functions of $c$, are in this case
\begin{equation}
E_{Bu}^{2}(\rho_{2})=Q_{Bu}^{2}(\rho_{2})= 2-\sqrt{2}\sqrt{1+\sqrt{1-c^2}},
\end{equation}
\begin{equation}
C_{Bu}^{2}(\rho_{2})=T_{Bu}^{2}(\rho_{2})=2-\sqrt{2}.
\end{equation}
In this case the non-additivity of correlations is evident, due to the fact that classical correlations and total correlations are identically equal even though quantum correlations are non-vanishing.
Furthermore, general quantum correlations of the discord type are equal to entanglement even though $\rho_{2}$ is not pure.

\section{Dynamics of Bures distance correlations}\label{sec:dynamics}
In this Section we study the dynamics of Bures distance correlations between two initially correlated but noninteracting qubits, each undergoing local pure dephasing due to the coupling with a bosonic bath at zero temperature and with super-Ohmic spectrum, whose characteristics do not depend on the qubit~\cite{Haikka2013}. Specifically, the dynamics of each qubit and the bosonic reservoir is governed by the following Hamiltonian ($\hbar=1$)
\begin{equation}
H=\omega_0\sigma_z + \sum_k\omega_k a_k^\dagger a_k + \sum_k \sigma_z \left(g_k a_k + g_k^*a_k^\dagger \right),
\end{equation}
where $\omega_0$ is the qubit frequency, $\omega_k$ the frequencies of the reservoir modes, $\sigma_z$ the Pauli operator along the $z$-direction, $a_k$ the bosonic annihilation operators, $a_k^\dagger$ the bosonic creation operators and $g_k$ the coupling constants between the qubit and each reservoir mode. Moreover, in the continuum limit we have $\sum_k {\left|g_k \right|}^2\rightarrow \int d\omega J(\omega)\delta(\omega_k - \omega)$, where $J(\omega)$ is the reservoir spectral density, which in the super-Ohmic case is given by
\begin{equation}
J(\omega)=\frac{\omega^s}{\omega_c^{s-1}}e^{-\omega/\omega_c},\ \ \ s>1
\end{equation}
with $\omega_c$ denoting the cut-off reservoir frequency.

If the two-qubit state is initially prepared in BD form with coefficients $c_1(0),c_2(0),c_3(0)$, then the ensuing dynamics preserves the BD form of the state, whose correlation coefficients evolve in time according to the following expressions,
\begin{eqnarray}
c_1(t)=q^2(t)c_1(0), \\
c_2(t)=q^2(t)c_2(0), \\
c_3(t)=c_3(0),
\end{eqnarray}
where the decay characteristic function $q(t)$ at zero temperature is given by $q(t)=e^{-\Upsilon(t)}$,
with dephasing factor
\begin{equation}
\Upsilon(t)=2\int_0^t\gamma(t')dt'
\end{equation}
and dephasing rate
\begin{equation}
\gamma(t)=\omega_c {\left[1+{\left(\omega_c t \right)}^2  \right] }^{-s/2}\Gamma[s]\sin\left[s \arctan(\omega_c t ) \right],
\end{equation}
where $\Gamma[x]$ is the Euler Gamma function.
We are interested in generally non-Markovian dynamics since it can prolong the existence of quantum correlations between the qubits. Consequently, in the following we will consider $s>s_{c}=2$, where $s_{c}=2$ represents the zero-temperature crossover between Markovian and non-Markovian dynamics~\cite{Haikka2013}.

\begin{figure}[t]
  \centering
  \subfigure[]{\includegraphics[width=0.6\textwidth]{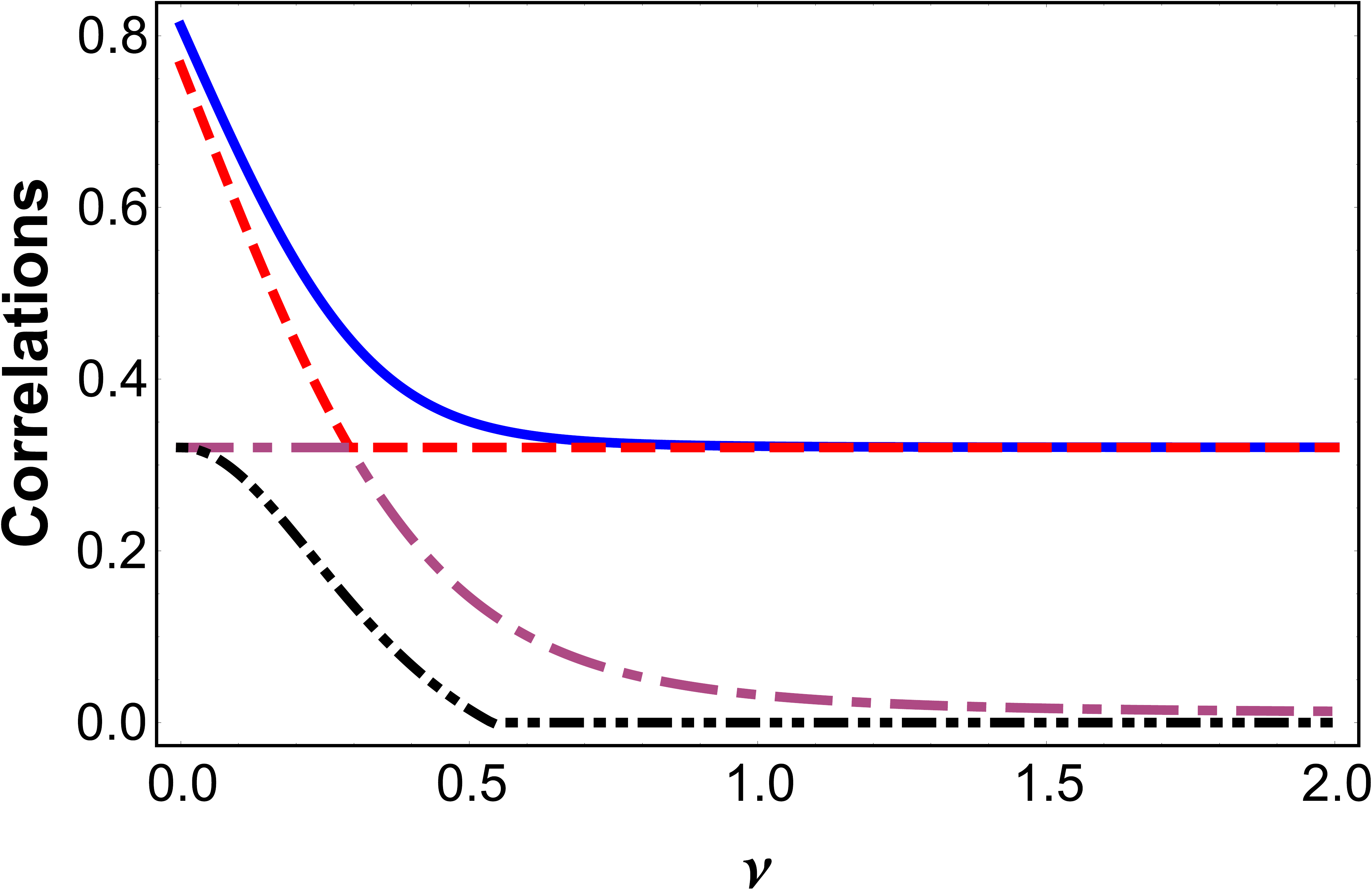}\label{Fig:Dynamics}} \\
  \subfigure[]{\includegraphics[width=0.6\textwidth]{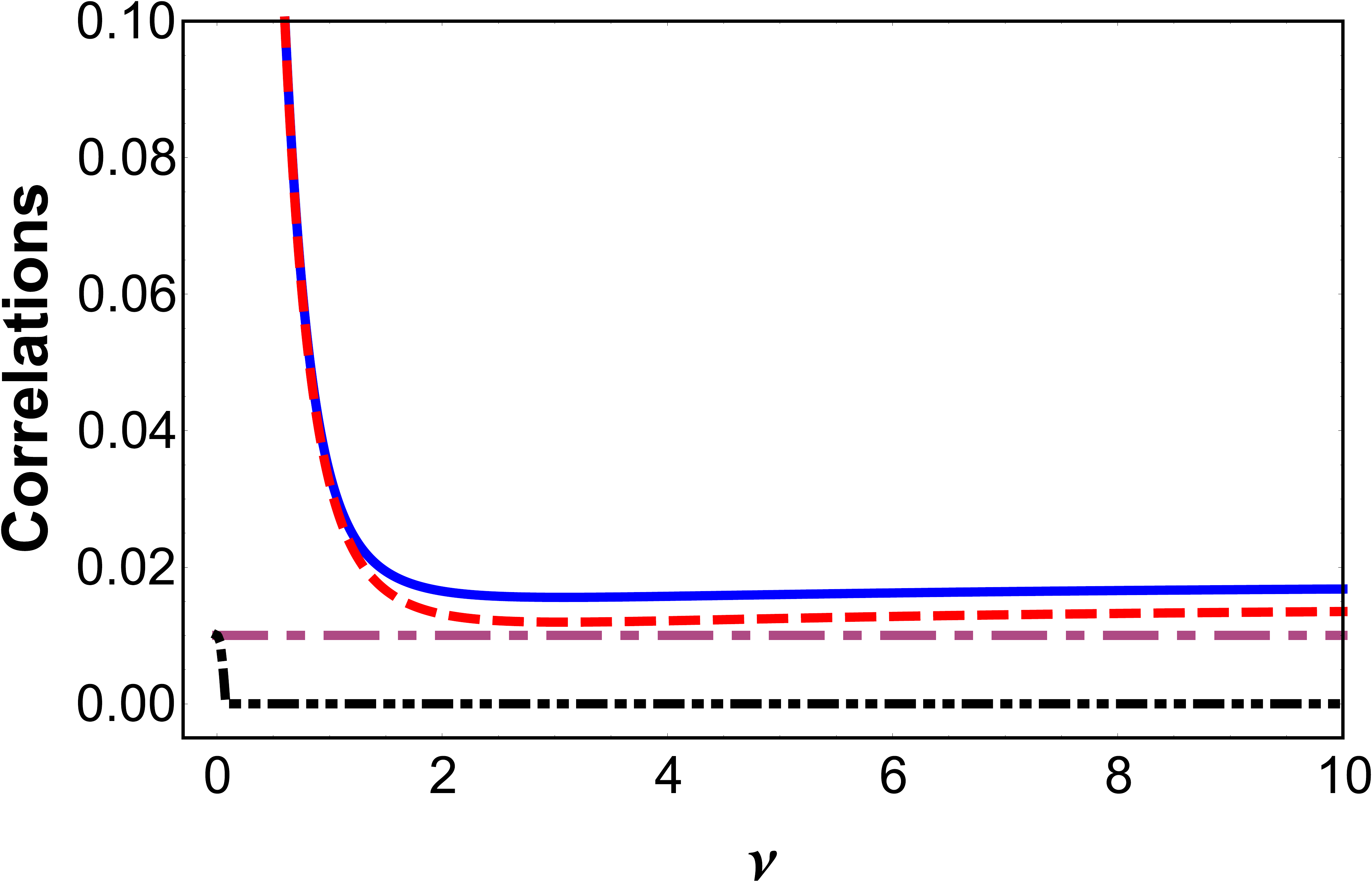}\label{Fig:Dynamics2}} \\
\caption{Dynamics of Bures distance correlations, plotted versus the dimensionless time  $\nu=\omega_c t$, for two noninteracting qubits, each coupled to a super-Ohmic bosonic reservoir with $s=2.5$ and initially prepared in a Bell-diagonal state with coefficients (a) $c_1(0)=1,c_2(0)=-0.6,c_3(0)=0.6$ and (b) $c_1(0)=1,c_2(0)=-0.02,c_3(0)=0.02$. The figures display total correlations (solid blue line), classical correlations (dashed red line), quantum correlations (dot-dashed purple line) and entanglement (dot-dot-dashed black line).}
    \label{Fig:DynamicsBoth}
\end{figure}

Figure \ref{Fig:Dynamics} shows the behaviour of Bures distance total, classical and quantum correlations as a function of the dimensionless time $\nu=\omega_c t$, when the two qubits are initially prepared in a rank-$2$ BD state with coefficients $c_1(0)=1$, $c_2(0)=-0.6$, $c_3(0)=0.6$ and each locally interacts with a zero-temperature super-Ohmic bosonic reservoir with $s=2.5$. This particular dynamics leads to paradigmatic {\it freezing} phenomena for both quantum and classical correlations, which can stay constant in certain time intervals despite the state physically changing; this was originally reported by analysing entropic measures of correlations \cite{Mazzola}. Bures measures of correlations are here shown to capture the same phenomenology. In our specific example, the freezing of quantum correlations can be seen up until a certain time $t^*$ and the freezing of classical correlations from that exact time. It is worth noting that classical and quantum correlations exactly coincide at time $t^*$, which is known as the instant of sudden transition from classical to quantum decoherence \cite{Mazzola}. This property also happens for correlations quantified by relative entropy \cite{Haikka2013,xulofranco2013NatComms,lofrancoreview} and Hilbert-Schmidt distance \cite{Bellomo2012a}. For correlations quantified by trace distance, although quantum and classical correlations alternately freeze in the same time intervals as the other measures \cite{Aaronson2013a,Aaronson2013,Paula2013}, they are generally not coincident at the crossing time $t^*$. The entanglement, instead, decays monotonically and completely disappears at a finite time exhibiting the well-known sudden death phenomenon \cite{lofrancoreview,yu2009Science}. Total correlations, finally, decay smoothly and asymptotically reach a non-vanishing value.

However, the aforementioned sudden switch between freezing of quantum and classical correlations  may not happen at all if the two qubits are initially prepared in a suitable BD state. This is displayed in Figure~\ref{Fig:Dynamics2} for initial coefficients $c_1(0)=1,c_2(0)=-0.02,c_3(0)=0.02$. In fact, in the latter case, quantum correlations are frozen indefinitely and never coincide with classical correlations. This peculiar time behaviour is in accordance with Ref.~\cite{Haikka2013}, in which this phenomenon was originally investigated by considering the entropic quantum discord \cite{Ollivier2001}. We also point out that, instead, the entanglement presents again a sudden death despite the indefinite freezing of general quantum correlations. This is an interesting point showing once more the very different nature of quantum correlations, as captured by the concept of quantum discord, and conventional entanglement: quantum correlations can be in principle completely unaffected by decoherence in spite of the fast disappearance of entanglement. For a more general study of the sudden change phenomena for quantum correlations see Ref.~\cite{FanchiniNew}.

\section{Conclusions}\label{sec:conclusions}
In this paper we obtained exact expressions for the classical and total correlations of two-qubit Bell-diagonal states according to the Bures distance. These expressions, together with the corresponding known formulae for entanglement and discord-type quantum correlations, allow for the completion of the hierarchy of Bures distance based correlations in the physically relevant family of Bell-diagonal states.   We proved that the closest Bell-diagonal classical-quantum state and the tensor product of the marginals of any Bell-diagonal state achieve the infima in the definition of classical correlations. We have shown that the nearest uncorrelated state to every Bell-diagonal state is not always the product of its marginals and that, in general, the total correlations are strictly smaller than the sum of quantum and classical correlations. These results are in agreement with those obtained with the trace distance \cite{Aaronson2013,PaulaEPL} but not with the Hilbert-Schmidt \cite{Bellomo2012a} and relative entropy distances \cite{Modi2010}. Finally, we have shown that, for two independent qubits under local bosonic pure-dephasing non-Markovian channels and suitable initial conditions \cite{Mazzola,Haikka2013}, Bures distance quantifiers of correlations can also manifest: (i) the freezing of quantum correlations before a certain time $t^*$ and the freezing of classical correlations after the same time $t^*$; (ii) the indefinite freezing of quantum correlations notwithstanding the contemporary occurrence of entanglement sudden death. Another peculiar feature of quantum but not classical geometric correlations, namely the possibility of double sudden changes (under  different dynamical conditions) as observed theoretically and experimentally for trace distance quantifiers \cite{SarandyNew,PaulaEPL,Paula2013}, has also been theoretically shown to occur for the Bures measure of discord \cite{OrszagNew}.

Overall, with the contributions brought forward by this paper, we have reached a fairly comprehensive understanding of the different components of correlations in generally mixed states of archetypical bipartite systems and their dynamical characteristics. Experimental demonstrations with quantum optics and nuclear magnetic resonance setups \cite{Guo,Cornelio2012,xulofranco2013NatComms,DiogoNMR,Isabela,Paula2013} have also verified the predicted resilience of classical and quantum correlations to particular decoherent evolutions. However, a satisfactory understanding of these phenomena from first principles is still missing. This is perhaps to be traced back to a still incomplete formalisation for the requirements that measures of classical and general quantum correlations have to satisfy \cite{Modi2012,Ollivier2001,Henderson2001}. Future work will be devoted to the theoretical and experimental understanding of the physical origin of the freezing under nondissipative decoherence common to any known \textit{bona fide} measure of quantum correlations before a certain time $t^*$ (including entropic, trace, Bures, and skew-information based measures as studied in \cite{Aaronson2013a}), and to any currently known \textit{bona fide} measure of classical correlations after the same time $t^*$ (counting entropic, trace and now Bures classical correlations). The ultimate implications of such extreme robustness to decoherence, not observable in entanglement, and the ensuing possibilities opened for quantum technologies clearly deserve further investigation.

\ack{
We thank Ben Aaronson, Massimo Blasone, Catalin Catana, Fabrizio Illuminati, Soojoon Lee, Giuseppe Marmo, Sammy Ragy,  Marcelo Sarandy, Diogo~O.~Soares-Pinto, Dominique Spehner and Karol \.{Z}yczkowski for fruitful discussions. We acknowledge the
Brazilian funding agency CAPES [Pesquisador Visitante Especial-Grant No.~108/2012] and the Foundational Questions Institute [FQXi-RFP3-1317] for financial support.}

\section*{References}

\providecommand{\newblock}{}

\end{document}